\documentstyle[amsfonts,aps]{revtex}

\title{Computation of Yukawa  Couplings for Calabi-Yau Hypersurfaces
        in Weighted Projective Spaces}
\preprint{CLNS 94/1295}
\author{ Yakov Kanter   }
\address{ F.R. Newman Laboratory of Nuclear Studies, Cornell University,
Ithaca, NY  14853 }

\begin{document}

\maketitle

\begin{abstract}

 Greene, Morrison and Plesser \cite{GMP} have recently suggested a
general method for constructing a mirror map between a $d$-dimensional
Calabi-Yau hypersurface and its mirror partner for $d > 3$. We
apply their method to smooth  hypersurfaces in
weighted projective spaces and compute the Chern numbers of
holomorphic curves on these hypersurfaces. As anticipated, the results
satisfy nontrivial integrality constraints. These examples differ from
those studied previously in that standard methods of
algebraic geometry which work in the ordinary projective space case
for low degree curves
are not  generally applicable. In the  limited special  cases in which
they do work we can get
 independent predictions, and we find agreement with our results.
\end{abstract}

\section{Introduction}
The mirror symmetry conjecture is increasingly being recognized as a
powerful computational tool.  One class of problems to which it has
been successfully applied is the calculation of Chern numbers $n_{j}$
of the parameter spaces of  holomorphic degree $j$ curves on certain special
Calabi-Yau  $d-$folds.  Conformal
field theory relates $n_{j}$'s in a unique way to integer coefficients
of certain $q-$expansions of Yukawa couplings on these Calabi-Yau
manifolds.  Mirror symmetry then allows one to calculate the Yukawa
couplings by reinterpreting them in terms of deformations of the Hodge
structure of the {\em mirror manifold}, thus effectively calculating
$n_{j}$ for all $j$. This was first done in \cite{CdGP} for the
quintic Calabi-Yau 3-fold and its mirror. This technology is
particularily  easy to apply
to smooth hypersurfaces in ${\bf P}^{d+1}$ or ${\bf WP}^{d+1}$ with
one-dimensional K\"{a}ler moduli spaces.  (A more difficult case of
multidimensional  moduli spaces is discussed in \cite{CdFKM1} \cite{CdFKM2}
\cite{HKTY1} \cite{HKTY2}).
Traditional methods of algebraic geometry can calculate  $n_{j}$'s for
$j \leq 3$ for projective space, giving exactly the numbers predicted
by  mirror symmetry.  This makes it desirable to extend the list of
Calabi-Yau manifolds for which Yukawa couplings have been explicitly
calculated. In \cite{GMP} this was accomplished for all smooth
CY-hypersurfaces in  ${\bf P}^{d+1}$ for $4 \leq d \leq 9$.  In this paper we
will compute Yukawa couplings for all hypersurfaces with Picard number
equal to one in weighted
projective spaces ${\bf WP}^{d+1}$  for $4 \leq d \leq 7$.  It should
be noted that while there is exactly one  CY-hypersurface of this type in
${\bf P}^{d+1}$ for every $d \geq 3$, there are usually several such
manifolds  for weighted projective spaces. Furthermore, these examples
are generally not accessible to traditional methods of algebraic
geometry even for low degree curves and hence they further emphasize
the power of physical methods.

\section{ Smooth Hypersurfaces in $WP^{d+1}$}
In this section we review very briefly the mirror symmetry conjecture
in the form proposed by the authors of \cite{GMP} and describe their algorithm
for computing Yukawa couplings as it applies to smooth hypersurfaces
in weighted projective spaces.  All of the material in this section is
quite standard (see e.g.  \cite{GMP} and reviews in \cite{Ess}).  It
is included here to make our presentation more self-contained.

\subsection{Some background on mirror symmetry}
Consider a superconformal Landau-Ginsburg model with superpotential
${\cal Q}$.  Take ${\cal Q}$ to be a quasi-homogeneous polynomial.  An
argument due to Greene, Vafa and Warner \cite{GVW}, Martinec \cite{Mart}
 and Witten
\cite{Wit2} shows that this theory is an ``analytic continuation'' to
 negative K\"{a}hler class of a $\sigma$-model with Calabi-Yau
manifold as a target space.  This manifold can be described as a
codimension one variety
\begin{equation}
{\cal W} = \{ x, {\cal Q}(x) = 0 \}
\end{equation}
in a weighted projective space. \\
\indent Truly marginal operators with $U(1)$ charge equal to $+1$ $(
-1 )$ in   the
superconformal theory correspond to complex structure (K\"{a}hler)
modes of the Calabi-Yau theory.  We will be also considering ``extended
K\"{a}hler moduli spaces'' $H^{p}(X,\Lambda^{p}T^{*})$ and ``extended
complex structure moduli spaces'' $H^{p}(X, \Lambda^{p}T)$.  Following
Witten \cite{Wit}, the couplings involving elements of $H^{p}(X,
\Lambda^{p}T^{*})$ will be called {\bf A}-model couplings, while
couplings involving elements of  $H^{p}(X,
\Lambda^{p}T)$ -- {\bf B}-model couplings. \\
\indent Now consider the action of the maximal group of scaling symmetries $G$
on both Landau-Ginsburg and Calabi-Yau models for ${\cal Q}$ of Fermat
type.   It is well known
(see, e.g. review \cite{GP} and references therein)  that orbifolding the
LG  theory by G simply simply changes the
sign of one of the $U(1)$ charges and gives a  theory isomorphic to the
original one.  For a CY model, dividing by G means taking the orbifold of
the target space.   The
unique connection with  a LG model now leads to the central result
of the mirror symmetry, namely, to the prediction that the
two resulting quantum $\sigma$-models are isomorphic (even though
they have different tree-level actions, since their target manifolds
will in general be topologically distinct).  In particular, all
$n$-point functions computed  in both $\sigma$-models must be equal
for corresponding modes. \\

This result can be used to calculate Chern numbers of holomorphic curves on
a variety ${\cal W}$ as first shown in \cite{CdGP}.  Consider a Yukawa
coupling  between elements of
$H^{p}({\cal W},\Lambda^{p}T^{*})$. It was shown  in \cite{StWi}
\cite{St} \cite{DSWW} \cite{CdGP}
 \cite{AspinMorris} that for the  case of a  1-dimensional
K\"{a}hler moduli space
\begin{equation}
\label{eYuk}
\langle O^{i}O^{j}O^{k} \rangle = \langle e_i \, e_j \, e_k \rangle +
 \sum_{l}\frac{q^l}{1-q^l} \cdot n_l ,
\end{equation}
where $O^j$ are operators corresponding to elements of
$H^{p}({\cal W},\Lambda^{p}T^{*})$,
 $q = e^{2 \, \pi \, i \, t}$,
 $t$ is a ${\Bbb Z}$-perodic local coordinate on the complexified
K\"{a}hler cone of ${\cal W}$ and $n_l$'s are the Chern numbers of
degree $l$ curves on ${\cal W}$. Mirror symmetry applied to
3-point functions allows one to interpret equation (\ref{eYuk}) as a
coupling between elements of  $H^{p}({\cal M},\Lambda^{p}T)$, where
${\cal M}$
is the mirror partner of ${\cal W}$.
In this interpretation $ \langle O^{i}O^{j}O^{k} \rangle$ is given by
the cup product
of elements of  $H^{p}({\cal M},\Lambda^{p}T)$ and is not corrected by
instantons  \cite{DG}.  Thus it can be computed using methods of
algebraic geometry as a function of $z$, where $z$ is some local
coordinate on the moduli space of complex structures of ${\cal M}$.
The link to equation (\ref{eYuk}) will be established once we know $t$
as a function of $z$.  More generally, mirror symmetry predicts that
there exists a one-to-one map between  $H^{p}({\cal M},\Lambda^{p}T)$
and  $H^{p}({\cal W},\Lambda^{p}T^{*})$ which we will call the {\em
generalized   mirror  map}, since unlike the 3-dimensional case
it  is {\em not} determined by $z \mapsto t(z)$.
The preceding discussion shows that our problem of calculating the
Chern numbers of the holomorphic curves on ${\cal W}$ can be reduced
to: \\
$\bullet$ Computing properly normalized Yukawa couplings on the mirror
partner of ${\cal W}$ \\
$\bullet$ Computing the generalized  mirror map. \\

The authors of \cite{GMP} suggested a solution \footnote{Apparently
their results can be justified using the methods of \cite{BCOV}}  to
both  problems by
conjecturing that the generalized  mirror map takes the primary
vertical  subspace
of $\bigoplus_{p}H^{p,p}({\cal W},{\Bbb Z})$ into the horizontal subspace
of the  Gauss-Manin connection on $\bigoplus_{p}H^{p}({\cal
W},\Lambda^{p}T)$.
Specifically, they showed that one can construct the mirror map in the
following way : \\

 1. Select a primary vertical subbasis $e_0, e_1, \cdots , e_d$ of
$\bigoplus_{p}H^{p,p}({\cal W},{\Bbb Z})$, satisfying the operator
product  equations
\begin{mathletters}
\label{eOpr}
\begin{equation}
\label{eOpr-a}
e_1 \cdot e_{j-1} = c^{-1} \, A_{j-1}(t) \, e_{j},
\end{equation}
\begin{equation}
 \langle e_i , e_j \rangle  = \eta^{(i,j)} = c \, \delta_{i+j,d},
\end{equation}
\begin{equation}
e_1 \cdot e_d = 0.
\end{equation}
\end{mathletters}

2. Declare that the mirror of this basis is a set of elements $\alpha_0$,
 $\alpha_1,
\cdots , \alpha_d$ of  $\bigoplus_{p}H^{p,d-p}({\cal M},{\Bbb C})$
with $\alpha_0 = \Omega \in H^{p,0}({\cal M}, {\Bbb C})$ such that the
same axioms hold for $\alpha_j$ with {\bf B}-model three-point
function $B_{j-1}(z)$ replacing $A_{j-1}(t)$ in equation
(\ref{eOpr-a}) and action of Gauss-Manin connection replacing operator
product
\begin{mathletters}
\label{eOprM}
\begin{equation}
\label{eOprM-a}
e_1 \cdot e_{j-1} = c^{-1} \, B_{j-1}(z) \, e_{j},
\end{equation}
\begin{equation}
 \langle e_i , e_j \rangle  = \eta^{(i,j)} = c \, \delta_{i+j,d},
\end{equation}
\begin{equation}
e_1 \cdot e_d = 0.
\end{equation}
\end{mathletters}

They further showed that conditions of (\ref{eOprM}) will be satisfied
for
$\alpha_j$'s which are   covariantly constant with respect to the
Gauss-Manin  connection  and  such that their period matrix
\begin{equation}
P_{\mu i} = \int_{\gamma_{\mu}} \alpha_{i}
\end{equation}
is upper-triangular with ones on the diagonal for homology cycles
$\gamma_0$, $\gamma_1, \cdots , \gamma_d$ satisfying the maximum
unipotency  condition at $z = 0$ (see equation (\ref{eq:defin-z}) for a
definition of $z$).
This   allows one to compute the mirror map and normalized Yukawa
couplings using the algorithm described in the next subsection.

\subsection{An Algorithm for Computing the Mirror Map}
Let ${\cal W}_{t}$ be a Fermat hypersurface in ${\bf WP}(k_1, \ldots ,
k_{d+1})$ defined by the vanishing loci of
\begin{equation}
Q(x) = x_1^{d_1} + \cdots + x_{d+1}^{d_{d+1}}
\end{equation}
with complex structure induced from $WP^{d}$ and a $t-$dependent
K\"{a}hler structure induced from
\begin{equation}
e(t) = t \cdot e ,
\end{equation}
where $e$ is some fixed generator of $H^{1,1}({\bf WP}^{d},{\Bbb Z})$.
Now consider
\begin{equation}
\tilde{{\cal M}}_{\psi} = \{ x \in {\bf WP}(k_1, \ldots
,k_{d_{d+1}}) \, | \,  Q(x,\psi) = 0 \},
\end{equation}
where
\begin{mathletters}
\label{eq:defin}
\begin{equation}
\label{eq:defin-Q}
Q(x,\psi) =  x_1^{d_1} + \cdots + x_{d+1}^{d_{d+1}} - k \psi \cdot x_1
\cdots x_{d+1},
\end{equation}
\begin{equation}
\label{eq:defin-k}
 k = \sum_{j} k_j,
\end{equation}
\begin{equation}
\label{eq:defin-z}
 z = \psi^{-k}.
\end{equation}
\end{mathletters}
 If  we choose the weights $k_j$ and degrees $d_j$ so that
$gcd(k_i,k_j)$=1 for $i \neq j$ , $d_j = k/k_j$, \\
then  ${\bf WP}(k_1, \ldots , k_{d+1})$ will be smooth away from the
origin and by Lefshetz's theorem $h^{1,1}(\tilde{{\cal M}}_{\psi}) = 1$ when
$\tilde{{\cal M}_{\psi}}$ is non-singular. \\
\indent Let ${\cal M}_{\psi} = \tilde{{\cal M}}_{\psi}/G$, where $G$
is the maximal group of scaling symmetries  \cite{GP}, with
$\psi-$dependent complex structure induced from ${\bf WP}^{d}$ and the
K\"{a}hler structure corresponding to the Landau-Ginsburg point in the
K\"{a}hler moduli space \cite{Wit2} \cite{AGM} .
Mirror symmetry now predicts that for every $\psi \in {\Bbb C}$
there exists $t(\psi) \in {\Bbb C}$ such that ${\cal M}_{\psi}$ is a
mirror of ${\cal W}_{t(\psi)}$, i.e. that quantum $\sigma-$models with
${\cal M}_{\psi}$ and ${\cal W}_{t(\psi)}$ as target manifolds are
isomorphic.  The map $\psi \mapsto t(\psi)$ together with Yukawa
couplings can be obtained as a part of the generalized mirror map in
the following way.

Step 1.\\
A form $\alpha \in H^{p,d-p}({\cal M}_{\psi}, {\Bbb C})$ is
covariantly constant with respect to Gauss-Manin connection if
\begin{equation}
\label{eq:period}
f(z) = \int_{\gamma(z)} \alpha
\end{equation}
satisfies a $d$-th order ordinary differential equation, called
 the Picard-Fuchs equation \cite{Arn}. An equivalent system of
first-order equations is
 \begin{equation}
\label{ePF}
 z \, \frac{d \, w}{d \, z} = A(z) \, w(z) ,
\end{equation}
where
\begin{equation}
w(z) = \left( \begin{array}{c}
        \int_{\gamma} \alpha  \\
         z \, \frac{d}{d \, z} \int_{\gamma} \alpha \\
          \vdots              \\
         (z \, \frac{d}{d \, z})^d \int_{\gamma} \alpha
              \end{array}
      \right),
\end{equation}

\begin{eqnarray}
 A(z) = \left( \begin{array}{cccc}
                  0        & 1      & \cdots    & 0 \\
                           & \ddots & \ddots    &  \\
                           &        &           & 1 \\
                  B_{0}(z) & \cdots &           & B_{d}(z)
               \end{array}
        \right)
\end{eqnarray}

The coefficients $B_0, \ldots , B_d$ encode all the information
necessary to compute Yukawa couplings for a specific hypersurface.
They can be generated using Griffiths' pole reduction method, which was
described in \cite{MorPic} for $d = 3$ but works for $d \ge 3$ without
modification.
  Just as in the $d=3$ case
the problem reduces to finding an explicit representation of polynomials
in a Jacobian ideal over ${\Bbb Q}(\psi)$.  This is  achieved in the
following way. \\
 Suppose we want to find $A_j$'s such that
\begin{equation}
\label{eq:repr}
   g = \sum_j A_j \cdot F_j,
\end{equation}
where $g$ and $F_j$'s are known polynomials.
 First,
we use the built-in ability of MACAULAY \cite{SB} to compute syzygies
to find $A_j$ over a finite field ${\Bbb F}_{31991} = {\Bbb Z}/31991 \cdot
{\Bbb Z}$.
In general, if ${\cal I}$ is an ideal with generators $f_1, \ldots ,
f_n$, then a syzygy of ${\cal I}$ is a $n$-dimensional vector $(g_1,
\cdots, g_n)$, such that $\sum_{i=1}^{n} g_{i} \cdot f_{i} = 0$. Thus
to find a representation (\ref{eq:repr}) we simply find a set of
generators for the syzygy module of ${\cal J} =  \langle g, F_1, \ldots
,F_n \rangle$ and  pick the one with a constant first element.
Then all
monomial coefficients of $A_j$ are  replaced with unknown constants and the
result substituted back into equation (\ref{eq:repr}).  The resulting
linear equations for these constants are then solved in MAPLE over
${\Bbb Q}(\psi)$, thus lifting the representation to ${\Bbb Q}(\psi)$.  \\
 \indent In all the examples we considered, $B{_j}$'s turned out to be
rational functions of  $z$,  as expected,  and
\begin{equation}
\label{eBs}
 B_0(0) = \ldots = B_d(0) =0
\end{equation}

The last condition ensures that $z = 0$ is a regular singular point of
(\ref{ePF}) and that the monodromy is maximally unipotent at $z = 0$.\\

Step 2.\\
A fundamental matrix of solutions of (\ref{ePF}) can be written in the form
\begin{eqnarray}
\label{eFund}
 \Phi (z) & = & S(z) \, z^{A(0)}   \nonumber  \\
          & = & S(z) \, \left( \begin{array}{cccc}
               1   &  \log (z)  &   & \frac{1}{n!} (\log (z))^{n}  \\
                   &  \ddots    & \ddots   &                             \\
                   &            &          & \log (z)   \\
                0  &            &          &  1
                              \end{array}  \right)
\end{eqnarray}
where $S(z)$ is a single-valued holomorphic matrix-valued function of $z$.
Substitution of (\ref{eFund})  into (\ref{ePF}) gives an equation for
$S$,
\begin{equation}
 z \, \frac{d S}{d z} + S(z) \cdot A(0) = A(z) \cdot S(z),
\end{equation}
 which  can  be solved using power series techniques.\\

Step 3.\\
Put the matrix $S$ in the upper triangular form using only row
operations.  Call the resulting $(d+1) \times (d+1)$ matrix $T$ and let
the indices of $T$ run from 0 to $d$.\\

The canonical parameter $t$ of \cite{MorGuide} is then given by
\begin{equation}
\label{et}
  t = T_{01}
\end{equation}
and the canonical variable of the $q-$expansion is, as usual,

\begin{equation}
\label{eexpt}
 q = \exp(2 \pi i \, t)
\end{equation}

Step 4.\\
Fundamental Yukawa couplings, i.e. Yukawa couplings involving at least
one element of $H^{-1,1}({\cal M}_{\psi})$ when interpreted as couplings
in {\bf B}-model of  \cite{Wit}, can be expressed in terms of the
first superdiagonal of $T$. Following \cite{GMP} we shall denote
$<O^{1} \, O^j \, O^{d-j-1}>$ as $Y_j^1$ .  Then it was shown in
\cite{GMP} that all couplings can be expressed in terms of the
fundamental ones using the associativity of the operator product
expansion and that
\begin{equation}
\label{eyuk}
 Y_j^1 = c \, \frac{1 + z \, \partial_z T_{j,j+1}}{1+z \,
\partial_z T_{0,1}},
\end{equation}
where $c$ is the degree of the surface, i.e. the smallest $d_j$ in
equation (\ref{eq:defin-Q}).

\section{Examples}
The authors of \cite{GMP}  compute Yukawa couplings for all smooth Picard
one Calabi-Yau hypersurfaces in ${\bf P}^{d+1}$  for $4 \leq d \leq
10$. There is exactly one such hypersurface for each
dimension $d$.  In this section we will compute the fundamental Yukawa
couplings for all Picard one hypersurfaces in ${\bf WP}^{d+1}$ for
$4 \leq d \leq 7$.

\subsection{Hypersurfaces in $WP^5$}
 There is exactly one set of $(k_1 , \ldots , k_6)$ that satisfies the
necessary conditions for the smoothness of ${\bf WP}(k_1, \ldots ,
k_6)$, namely (5,1,1,1,1,1).
 Thus the only family of Calabi-Yau submanifolds is given by
\begin{equation}
\label{eWP5}
 Q = x_1^2 + x_2^{10} + \cdots + x_6^{10} - 10 \, \psi \, x_1 \cdots
x_6 = 0
\end{equation}

There is only one independent 3-point function $Y_1^1$ in this case.
We find
\begin{equation}
  Y_1^1 = 2+1582400\frac{q}{1-q}+3167779945600\frac{q^2}{1-q^2}
+7052557179599697600\frac{q^3}{1-q^3}+ \ldots
\end{equation}

\subsection{Hypersurfaces in $WP^6$ -- $WP^{8}$}
All smooth hypersurfaces in $WP^{d+1}$ $5 \leq d \leq 7$  are
described in  tables
\ref{table:dim5} -- \ref{table:dim7}.
The fundamental Yukawa couplings are summarized in tables
\ref{table:YukDim5} -- \ref{table:YukDim6}.
Methods of algebraic geometry allow one to compute Yukawa couplings
for weighted projective spaces with all but one weight equal to 1. In
particular, Sheldon Katz has confirmed our result for $Y_1^2$ in ${\bf
WP}^{6}(2,1,1,1,1,1)$ \cite{katz}

\section{Conclusions}
In this paper we used the method described in \cite{GMP} to  calculate
 normalized Yukawa couplings for
Calabi-Yau hypersurfaces of complex dimensions $d = 4, \ldots ,  7$ in weighted
projective spaces. The fact that the coefficients of Yukawa couplings'
expansions in canonical variable $q$  turn out to be integers satisfying
rather intricate divisibility constraints described in \cite{GMP}
provides  support for the
mirror symmetry conjecture in this case. We have found agreement
between  our results and those of more standard mathematical
techniques when the latter are applicable. Finally, the general case of
a Calabi-Yau hypersurface in weighted projective space is not
accessible to standard methods of algebraic geometry and hence extends
the domain of cases which can only be analyzed with physical methods.

\section{Acknowledgments}
It is a pleasure to acknowledge very helpful discussions with Bill Dimm,
Brian Greene, Mike Stillman and Bernd Strumfels. This work was
partially supported by a grant from the  National Science Foundation.

\begin{table}
\caption{Smooth hypersurfaces of 6-dimensional weighted projective spaces}
\label{table:dim5}
\begin{tabular}{|l|l|l|}
$(k_1, \cdots k_7)$ & $k=\sum_{j=1}^7 k_j$ & Defining equation in
$WP^6$ \\ \hline
(2,1,1,1,1,1,1) & 8 &  $x_1^4 + x_2^{8} + \cdots + x_7^{8} - 8 \,
\psi \, x_1 \cdots x_7 = 0$ \\ \hline
(3,1,1,1,1,1,1) & 9 & $x_1^3 + x_2^{9} + \cdots + x_7^{9} - 9 \,
\psi \, x_1 \cdots x_7 = 0$ \\ \hline
(6,1,1,1,1,1,1) & 12 &  $x_1^2 + x_2^{12} + \cdots + x_7^{12} - 12 \,
\psi \, x_1 \cdots x_7 = 0$ \\ \hline
(4,3,1,1,1,1,1) & 12 &  $x_1^3 + x_2^{4} + x_3^{12} + \cdots + x_7^{12} - 12 \,
\psi \, x_1 \cdots x_7 = 0$ \\ \hline
(7,2,1,1,1,1,1) & 14 &  $x_1^2 + x_2^{7} + x_3^{14} + \cdots + x_7^{14} - 14 \,
\psi \, x_1 \cdots x_7 = 0$ \\
\end{tabular}
\end{table}

\begin{table}
\caption{Smooth hypersurfaces of 7-dimensional weighted projective spaces}
\label{table:dim6}
\begin{tabular}{|l|l|l|}
$(k_1, \cdots k_8)$ & $k=\sum_{j=1}^8 k_j$ & Defining equation in
$WP^7$ \\ \hline

(7,1,1,1,1,1,1,1) & 14 &  $x_1^2 + x_2^{14} + x_3^{14} + \cdots + x_8^{14} -
 14 \, \psi \, x_1 \cdots x_8 = 0$ \\
\end{tabular}
\end{table}

\begin{table}
\caption{Smooth hypersurfaces of 8-dimensional weighted projective spaces}
\label{table:dim7}
\begin{tabular}{|l|l|l|}
$(k_1, \cdots k_9)$ & $k=\sum_{j=1}^9 k_j$ & Defining equation in
$WP^8$ \\ \hline
(2,1,1,1,1,1,1,1,1) & 10 &  $x_1^5 + x_2^{10} + \cdots + x_9^{10} - 10 \,
\psi \, x_1 \cdots x_9 = 0$ \\ \hline
(4,1,1,1,1,1,1,1,1) & 12 & $x_1^3 + x_2^{12} + \cdots + x_9^{12} - 12 \,
\psi \, x_1 \cdots x_9 = 0$ \\ \hline
(5,3,1,1,1,1,1,1,1) & 15 &  $x_1^3 + x_2^{5} + x_3^{15} + \cdots + x_9^{15} -
 15 \,
\psi \, x_1 \cdots x_9 = 0$ \\ \hline
(8,1,1,1,1,1,1,1,1) & 16 &  $x_1^2 + x_2^{16} + x_3^{16} + \cdots + x_9^{16} -
 16 \,
\psi \, x_1 \cdots x_9 = 0$ \\ \hline
(9,2,1,1,1,1,1,1,1) & 18 &  $x_1^2 + x_2^{9} + x_3^{18} + \cdots + x_9^{18} -
 18 \,
\psi \, x_1 \cdots x_9 = 0$ \\
\end{tabular}
\end{table}

\begin{table}
\caption{Fundamental 3-point functions for smooth hypersurfaces in
$WP^6$}
\label{table:YukDim5}
\begin{tabular}{|l|l|}
(2,1,1,1,1,1,1) & $Y_1^1 = 4 + 3049216\frac{q}{1-q} + 7472581386752\frac{q^2}
{1-q^2} +
   21454661245363681536\frac{q^3}{1-q^3} + \ldots$ \\
                & $Y_2^1 = 4 + 5379584\frac{q}{1-q} + 16429968863232\frac{q^2}
{1-q^2} +
   55188836029204818432\frac{q^3}{1-q^3} + \ldots$  \\ \hline
(3,1,1,1,1,1,1) & $Y_1^1 = 3 + 8022402\frac{q}{1-q} + 66933014780124\frac{q^2}
{1-q^2} +
   657680002962846783606\frac{q^3}{1-q^3} + \ldots$ \\
                & $Y_2^1 = 3 + 13973877\frac{q}{1-q} + 146713683733290
\frac{q^2}{1-q^2}
+ 1692469099684864365660\frac{q^3}{1-q^3} + \ldots$  \\ \hline
(6,1,1,1,1,1,1) & $Y_1^1 =
2+71754624\frac{q}{1-q}+7975093545660672\frac{q^2}{1-q^2}+$ \\
                & $1044039441585542582459520\frac{q^3}{1-q^3} + \ldots$ \\
                & $Y_2^1 =
2+126008064\frac{q}{1-q}+17658771986147328\frac{q^2}{1-q^2}+$ \\
                & $2719158056221778746705152\frac{q^3}{1-q^3} + \ldots$
\\ \hline
(4,3,1,1,1,1,1) & $Y_1^1 =
3+765516096\frac{q}{1-q}+568850164748055936\frac{q^2}{1-q^2}+$ \\
                & $504680055676890191453456064\frac{q^3}{1-q^3} + \ldots$ \\
                & $Y_2^1 = 3+1400499072\frac{q}{1-q}+
1338726917072077056\frac{q^2}{1-q^2}+$ \\
                & $1411413775244683001901233280\frac{q^3}{1-q^3} + \ldots$
\\  \hline
(7,2,1,1,1,1,1) & $Y_1^1 =
2+1301207936\frac{q}{1-q}+2541843733963905280\frac{q^2}{1-q^2} +$ \\
                & $5895640558847778162251490432\frac{q^3}{1-q^3} + \ldots$ \\
                & $Y_2^1 = 2+2434050304\frac{q}{1-q}+
6058326351439047168\frac{q^2}{1-q^2}+$ \\
                & $16648147547178442866316220160\frac{q^3}{1-q^3} + \ldots$
\\
\end{tabular}
\end{table}

\begin{table}
\caption{Fundamental 3-point functions for smooth hypersurfaces in
$WP^7$}
\label{table:YukDim6}
\begin{tabular}{|l|l|}
(7,1,1,1,1,1,1,1) & $Y_1^1 = 2 + 3237982720 \frac{q}{1-q} +
21183078150223087616 \frac{q^2}{1-q^2} +$ \\
  &  $175301491479186058292989251072 \frac{q^3}{1-q^3} + \ldots$ \\
                & $Y_2^1 = 2 + 8106083328 \frac{q}{1-q} +
78167476562234465280 \frac{q^2}{1-q^2} +$ \\
  & $846729903166068966083368713216 \frac{q^3}{1-q^3} + \ldots$  \\
\end{tabular}
\end{table}

\begin{table}
\caption{Fundamental 3-point functions for smooth hypersurfaces in
$WP^8$}
\label{table:YukDim7}
\begin{tabular}{|l|l|}
2,1,1, \ldots 1 &
 $Y_1^1 = 5 + 873342000 \frac{q}{1 - q}
+ 850904087051808000 \frac{q^2}{1 - q^2} +$ \\
 & $1126679621495881973966658000 \frac{q^3}{1 - q^3} + \ldots$ \\

  & $Y_2^1 = 5 + 2788544000 \frac{q}{1 - q} + 4391772046773856000
\frac{q^2}{1 - q^2} +$ \\
   &  $ 8088656001785886785691456000
 \frac{q^3}{1 - q^3} + \ldots$ \\

 & $Y_3^1 := 5 + 3996616000 \frac{q}{1 - q}
+ 7539434669357184000 \frac{q^2}{1 - q^2} +$ \\
  &  $15863782567300449640323384000 \frac {q^3}{1 - q^3} + \ldots$ \\ \hline

4,1,1, \ldots 1 & $Y_1^1 = 3+7258320576
\frac{q}{1-q}+96985196693399291904 \frac{q^2}{1-q^2} +$ \\
  & $1777393929237545056819269167424 \frac{q^3}{1-q^3} + \ldots$ \\
  & $Y_2^1 = 3+23593099392 \frac{q}{1-q}+516294676633310908416
\frac{q^2}{1-q^2}+$ \\
 & $13238896229337709463174524669824 \frac{q^3}{1-q^3} + \ldots$
\\
 & $Y_3^1 =  3+33901042752\frac{q}{1-q}+892907832989728097280
\frac{q^2}{1-q^2} +$ \\
  & $26228563889472641235997337881536 \frac{q^3}{1-q^3} +
\ldots$ \\ \hline

8,1,1, \ldots 1
   & $Y_1^1 = 2+152056940544 \frac{q}{1-q}+63766419662687527673856
\frac{q^2}{1-q^2}+$ \\
 & $36785354090785907492391715484153856 \frac{q^3}{1-q^3} + \ldots$  \\
  & $Y_2^1 = 2+498854371328 \frac{q}{1-q}+
344276495484191699451904 \frac{q^2}{1-q^2}+$ \\
 & $278600071980095768674546472262721536 \frac{q^3}{
1-q^3} + \ldots$  \\
  & $Y_3^1 = 2+718234344448 \frac{q}{1-q}+597762641310376010833920
\frac{q^2}{1-q^2}+$ \\
 & $554788665268211922034998620987433984 \frac{q^3}{1-q^3} + \ldots$  \\
 \hline
5,3,1, \ldots 1
   & $Y_1^1 = 3+1004767067700 \frac{q}{1-q}+1935016728809768017084800
\frac{q^2}{1-q^2}$ \\
 & $+5206880462500493951270650091011932300 \frac{q^3}{1-q^3} + \ldots$  \\
  & $Y_2^1 = 3+3609099738900 \frac{q}{1-q}+
11658783428685286478013600 \frac{q^2}{1-q^2}$ \\
     & $+44401976701844742556180178323097636100 \frac{q
^3}{1-q^3} + \ldots$  \\
  & $Y_3^1 = 3+5344739516475 \frac{q}{1-q}+21077586779051709641435400
\frac{q^2}{1-q^2}$ \\
  & $+92729708830552378723454519048709905400 \frac{q^3}{1-q^3} + \ldots$  \\
\hline

9,2,1, \ldots 1
   & $Y_1^1 = 2+3306710034432 \frac{q}{1-q}+31348838504534376710504448
\frac{q^2}{1-q^2}$ \\
  & $+ 413657258502381181112268855777401837568 \frac{q^3}{1-q^3} + \ldots$  \\
  & $Y_2^1 = 2+11682851512320 \frac{q}{1-q}+
185001959444922156143738880 \frac{q^2}{1-q^2}$ \\
  & $+3448977159195424871434590786027577221120 \frac{q^3}{1-q^3} + \ldots$  \\
  & $Y_3^1 = 2+17375270943744 \frac{q}{1-q}+
334695839710556882880626688 \frac{q^2}{1-q^2}$ \\
  & $+ 7196484971475862667134042111055293694976 \frac{q^3}{1-q^3} +
 \ldots$  \\
\end{tabular}
\end{table}

\end{document}